\documentclass[11pt,letterpaper]{article}
\usepackage{systeme} 
\usepackage[utf8]{inputenc}
\usepackage{multirow} 
\usepackage{booktabs} 
\usepackage[english]{babel}
\usepackage{amsmath}
\usepackage{amsfonts}
\usepackage{amssymb}
\usepackage{graphicx}
\usepackage[small,bf]{caption} 
\usepackage[left=3cm,right=3cm,top=2cm,bottom=3cm]{geometry}

\author{Jorge Pinochet}
\title{\textbf{Classical Tests of General Relativity Part I: Looking to the Past to Understand the Present
}}
\begin{document}

\author{Jorge Pinochet$^{*}$\\ \\
 \small{$^{*}$\textit{Departamento de Física, Universidad Metropolitana de Ciencias de la Educación,}}\\
 \small{\textit{Av. José Pedro Alessandri 774, Ñuñoa, Santiago, Chile.}}\\
 \small{e-mail: jorge.pinochet@umce.cl}\\}

\date{}
\maketitle

\begin{center}\rule{0.9\textwidth}{0.1mm} \end{center}
\begin{abstract}
\noindent Einstein's theory of general relativity (GR) provides the best available description of gravity. The recent detection of gravitational waves and the first picture of a black hole have provided spectacular confirmations of GR, as well as arousing substantial interest in topics related to gravitation. However, to understand present and future discoveries, it is convenient to look to the past, to the classical tests of GR, namely, the deflection of light by the Sun, the perihelion precession of Mercury, and the gravitational redshift of light. The objective of this work is to offer a non-technical introduction to the classical tests of GR. In this first part of the work, some basic concepts of relativity are introduced and the principle of equivalence is analysed. The second part of the article examines the classical tests. \\ \\


\begin{center}\rule{0.9\textwidth}{0.1mm} \end{center}
\end{abstract}

\maketitle

\section{Introduction}
The theory of general relativity (GR), proposed by Einstein in 1915 [1,2], is the best available description of gravity. The first detection of gravitational waves in 2015 [3] and the first picture of a black hole obtained in 2019 [4–9] have provided spectacular confirmations of Einstein's theory, arousing significant interest among the general public in topics related to gravitation. There is no doubt that in the next few years, we will witness new confirmations of GR, but to understand the present and future discoveries, it is convenient to look to the past, to the classical tests of GR, namely, the deflection of light by the Sun, the perihelion precession of Mercury, and the gravitational redshift of light\footnote{A comment for experts: Strictly speaking, the gravitational redshift of light is a test of the Einstein equivalence principle, while the other two are tests of the GR in the weak field limit.} [10]. These tests were proposed by Einstein between 1907 and 1916 [1,2,11] and led to the first empirical confirmations of GR.\\

GR is a fascinating but highly technical subject, with its detailed understanding requiring advanced mathematical knowledge. The objective of this work is to offer a non-technical introduction to the classical tests of GR using simple mathematics. Therefore, the article can be used, for example, as educational material in an undergraduate modern physics course or in an introductory astronomy course.\\

In this first part of the work, some basic concepts of relativistic physics are presented as an introduction to the detailed analysis of the classical tests, which we will address in Part II. Section 2 discusses the weak equivalence principle, while Section 3 addresses Einstein equivalence principle, which is the foundation of GR. Section 4 outlines a qualitative explanation of the concept of spacetime curvature, and discusses its relationship to the equivalence principle. The article ends with a few brief summarising comments.

\section{Weak equivalence principle}

This principle has been known since the time of Galileo, but no one until Einstein was able to grasp its profound physical significance. The weak equivalence principle can be formulated as follows: \textit{the motion of any free falling test particle \footnote{A test particle is an ideal object that has such a small mass that its gravity can be ignored and therefore does not affect the gravitational field in which it is immersed.} is independent of its composition and structure}. To unravel the meaning of this principle, let us start by remembering that according to Newton's law of gravitation, the magnitude of the attractive force $F_{grav}$ generated by a spherical body of gravitational mass $M_{g}$ on a test particle of gravitational mass $m_{g}$ is:

\begin{equation}
F_{grav} = \frac{GM_{g}m_{g}}{r^{2}},
\end{equation}

where $r$ is the distance that separates the particle from the centre of the spherical body. On the other hand, according to Newton’s second law, the net force acting on a particle of inertial mass $m_{i}$, is related to acceleration as\footnote{Inertial mass is a measure of the resistance a body offers to be accelerated; for a given force, the greater the mass, the lower the acceleration. Instead, the gravitational mass is a measure of the attraction experienced by two bodies separated by a certain distance; the greater the mass, the greater the attraction.}:

\begin{equation}
F_{net} = m_{i}a.
\end{equation}

If we assume that $F_{grav} = F_{net}$, we obtain: 

\begin{equation}
a = \left( \frac{m_{g}}{m_{i}} \right) \frac{GM_{g}}{r^{2}}. 
\end{equation}

Numerous experiments have confirmed that $m_{g} = m_{i}$, and therefore: 

\begin{equation}
a = \frac{GM_{g}}{r^{2}}. 
\end{equation}

We see that all particles fall with the same acceleration, and consequently, the movement of the test particles is independent of their composition and structure, as stated by the principle of weak equivalence. Although this result may seem natural to us, to the point that many students and teachers automatically apply it in their calculations, it is important to note that other forces do not satisfy it. The simplest example is Coulomb law. According to this law, the magnitude of the electrostatic force between a body with charge $Q$ and a test charge $q$ is:

\begin{equation}
F_{elec} = \frac{KQq}{r^{2}},
\end{equation}

where $r$ is the distance between the charges. Let $m_{i}$ be the inertial mass of the test charge. If $F_{elec}$ is the only force acting on $q$, according to Eq. (2), we will have:

\begin{equation}
a = \left( \frac{q}{m_{i}} \right) \frac{KQ}{r^{2}}.
\end{equation}

Unlike Eq. (4), in this expression the acceleration depends on the $q/m_{i}$ ratio, that is, it depends on the composition and structure of the particles involved.\\

In the weak equivalence principle, expressed through Eq. (4), Einstein discovered something fundamental: if the movement of the test particle is independent of its composition and structure, then said movement is determined by a property that resides solely in the gravitational field. This property is the curvature of spacetime, since in relativity, space and time are inextricably linked, and what happens with spatial coordinates determines what happens with time intervals and vice versa.\\

Curvature is the central concept of GR and it conceives of gravity as a purely geometric property of the spacetime structure, where the mass induces curvature. That is, in the Einstein universe, the force of gravity does not exist: what moves planets, stars and galaxies is the curvature of spacetime.

\section{Theory of relativity and Einstein equivalence principle}

There are two formulations of the theory of relativity. The first, known as special relativity, was published by Einstein in 1905 and applies to phenomena where gravity is absent or so small that it can be ignored [2,12]. The second formulation, GR, was published by Einstein in 1916, and is a generalisation of special relativity that proposes a revolutionary description of gravity, understood as a purely geometric phenomenon [1,2]. The foundation of GR is the \textit{Einstein equivalence principle}: \textit{Experiments made locally in a reference frame uniformly accelerating with acceleration $\vec{a}$ whit respect to an inertial frame, produces the exact same experimental results as an inertial frame of reference in a uniform gravitational field $-\vec{a}$}[13,14].\\

We will use an example to clarify the meaning of the Einstein equivalence principle. Let us imagine a spacecraft traveling with constant acceleration $g$, in a region of the universe without gravity (see Fig. 1). An astronaut $A$ inside the ship has mass $m$ and is standing on a scale. What value does the scale record? Since the only force acting on $A$ is the normal $N$ exerted by the scale in the direction of $g$, it follows from Newton’s second law that the recorded value is $N = mg$.\\

If we consider the same spacecraft at rest on the surface of a planet where the gravitational acceleration is $–g$, the scale will also record the value $mg$. Furthermore, it can be shown that no experiment carried out inside the spacecraft will allow $A$ to determine if it is at rest on the surface of a planet with gravity $–g$, or if it is in a place in the universe without gravity moving with uniform acceleration $g$ [15]. However, this conclusion is only valid locally, since the gravity of a celestial body is not uniform, it varies with position and height. The gravitational effects can then only be reproduced locally, in small regions of space where gravity can be considered uniform.\\

\begin{figure}
  \centering
    \includegraphics[width=0.7\textwidth]{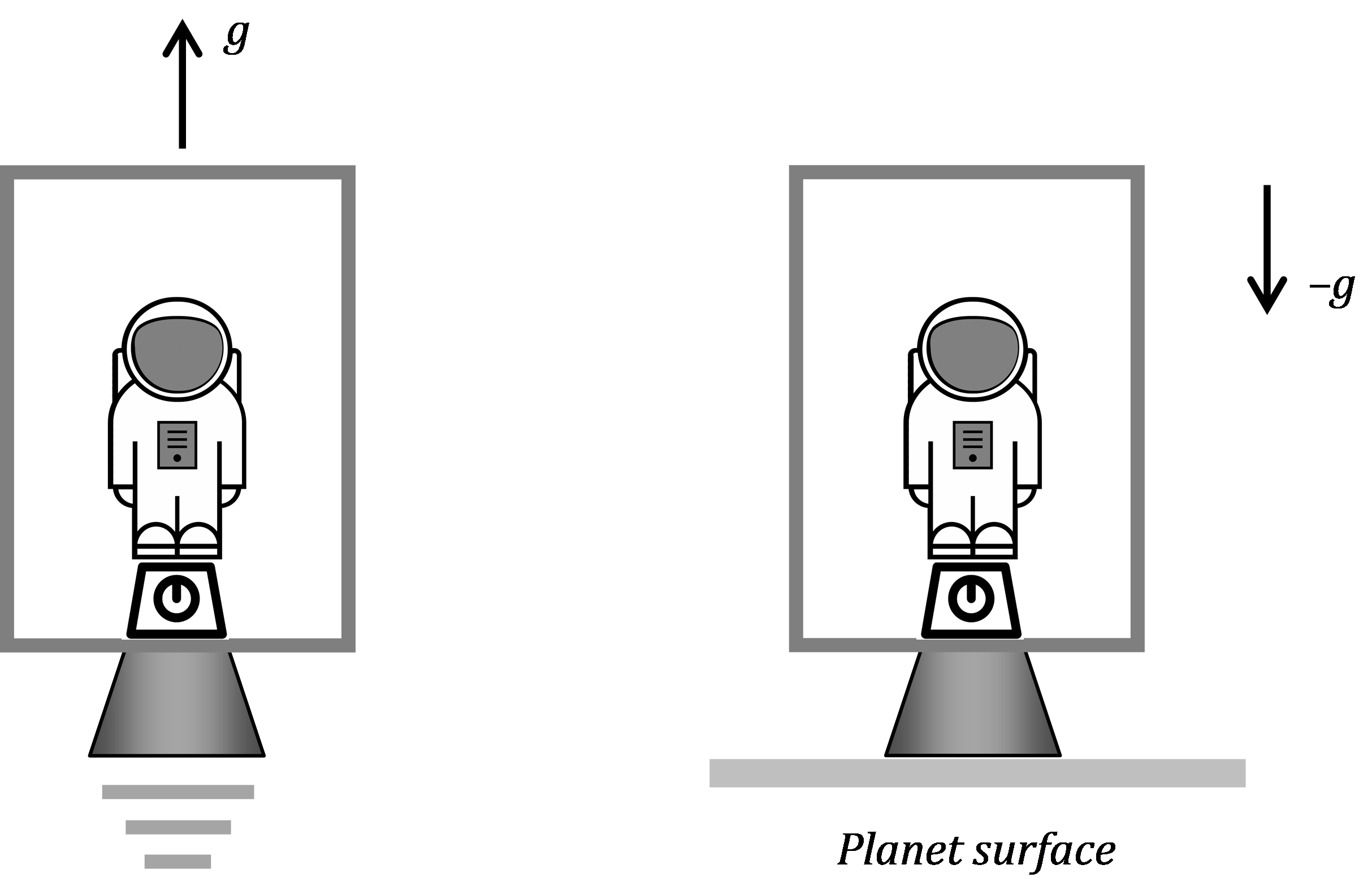}
  \caption{Left: An astronaut is standing on a scale inside a spacecraft traveling with gravitational acceleration $g$ through space without gravity. Right: The astronaut is at rest on the surface of a planet with a surface acceleration of gravity $–g$. In both cases, the scale shows $mg$.}
\end{figure}

However, the Einstein equivalence principle not only allows the local effects of gravity to be reproduced, but also allows us to eliminate these effects locally for a free falling reference frame under the action of a gravitational field. The latter leads to an alternative formulation of the Einstein equivalence principle: \textit{A free-falling reference frame in a gravitational field is locally equivalent to an inertial frame (without gravity)} [13,14].\\

To understand this new formulation, let us go back to the image of the spacecraft that we will now suppose encounters with its engines off and free falling toward a massive celestial body. Let us imagine that $A$ has a portable video camera. As the spacecraft falls, $A$ activates the camera and releases it from his or her hands (see Fig. 2). If the camera is pointed at $A$, what will it record? An image will be seen where $A$ floats motionless and stationary, as if in empty space, far from any source of gravity. Furthermore, no experiment carried out inside the spacecraft will allow $A$ to determine if it is in free falling under the action of the celestial body's gravity, or if it is in some remote corner of the universe, far from any source of gravity. However, again, due to the non-uniformity of the gravity field, the above conclusions are only valid locally in comparatively small regions.\\

\begin{figure}
  \centering
    \includegraphics[width=0.2\textwidth]{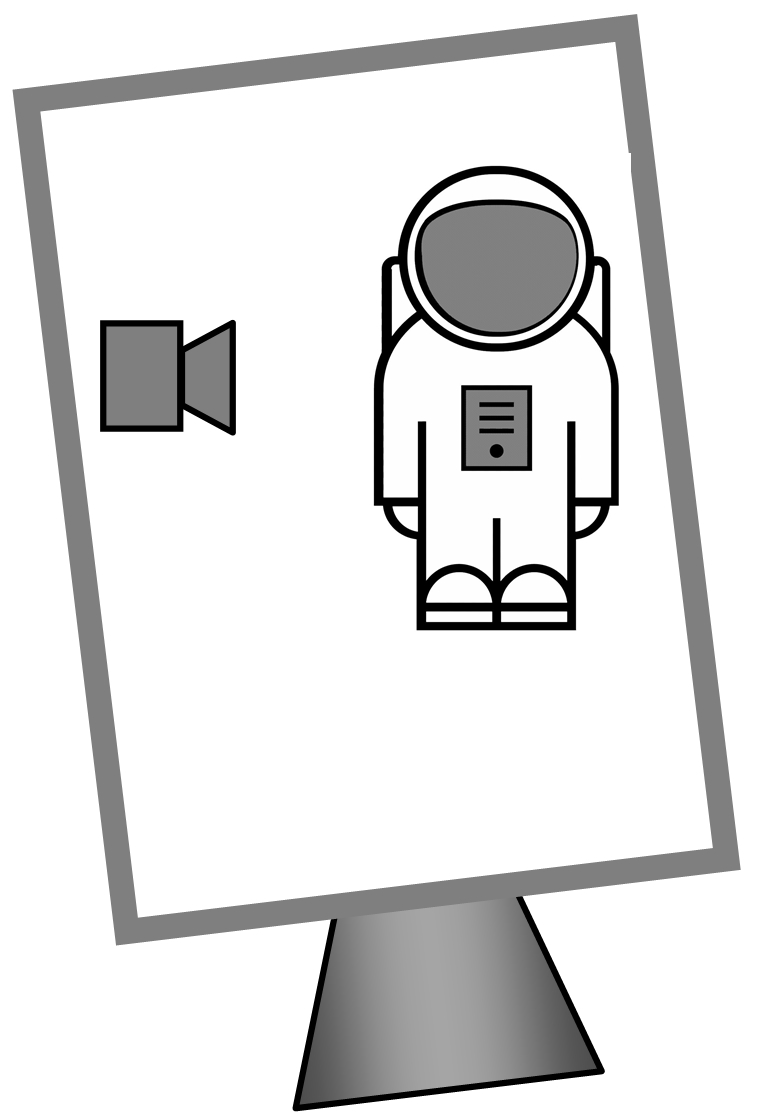}
  \caption{A free-falling astronaut appears to float inside the spacecraft, as if there is no gravity.}
\end{figure}

The Einstein equivalence principle is based on the weak equivalence principle. Indeed, if the movement of a body in a gravitational field depended on its composition and structure, as happens with electrical charges, we would see that the camera and the astronaut in the last example would fall differently, and we could not locally cancel the effect of gravity for a free falling reference frame. As noted in the previous section, it was this property of gravity that led Einstein to the conclusion that the motion of an object in a gravitational field depends on a property that resides solely in the field, and that corresponds to the spacetime curvature.

\section{Curvature of spacetime and Einstein equivalence principle}

Although the reading of this section is not strictly necessary to understand the ideas developed in Part II of this work, this section complements and enriches some of the topics developed before.\\

The central idea of GR is that gravity is a manifestation of the curvature of spacetime. Unlike in Newtonian physics, in GR, space and time are dynamic and flexible entities that respond to the presence of mass, or its energy equivalent. Thus, mass-energy curves spacetime and said curvature determines the movement of bodies. Therefore, in GR, there is no force of gravity. Bodies move freely following the trajectories dictated by the curved geometry of spacetime. Due to the curvature, these trajectories are not straight but \textit{geodetic}. A geodesic generalises the notion of the straight line of the flat Euclidean geometry; that is, a geodesic is a curve representing the shortest path between two points.\\

The American theoretical physicist John Archibald Wheeler beautifully described the idea behind GR, noting that "mass tells spacetime how to curve, and spacetime tells mass how to move" [16]. To understand the basic ideas of GR, it is useful to use some analogies with curved surfaces. In general terms, a two-dimensional surface can have only three kinds of curvature: \textit{null}, \textit{positive} and \textit{negative curvature}, as illustrated in Fig. 3.\\

The best example of zero curvature is the surface of a stretched sheet of paper. Imagine two particles that can only move on the surface of the sheet. For particles, the paper surface is their entire universe. On this surface, the classic results of flat Euclidean geometry are fulfilled. For example, (R1) \textit{if the particles move following parallel (straight) geodesics, they will keep their mutual distance constant (their trajectories do not intersect)} (see Fig. 3, left). It is usual to take this result as a criterion to define the zero curvature\footnote{Another criterion to define the zero curvature is the Euclidean theorem that states that the sum of the interior angles of a triangle is $180^{o}$}. \\

\begin{figure}
  \centering
    \includegraphics[width=0.7\textwidth]{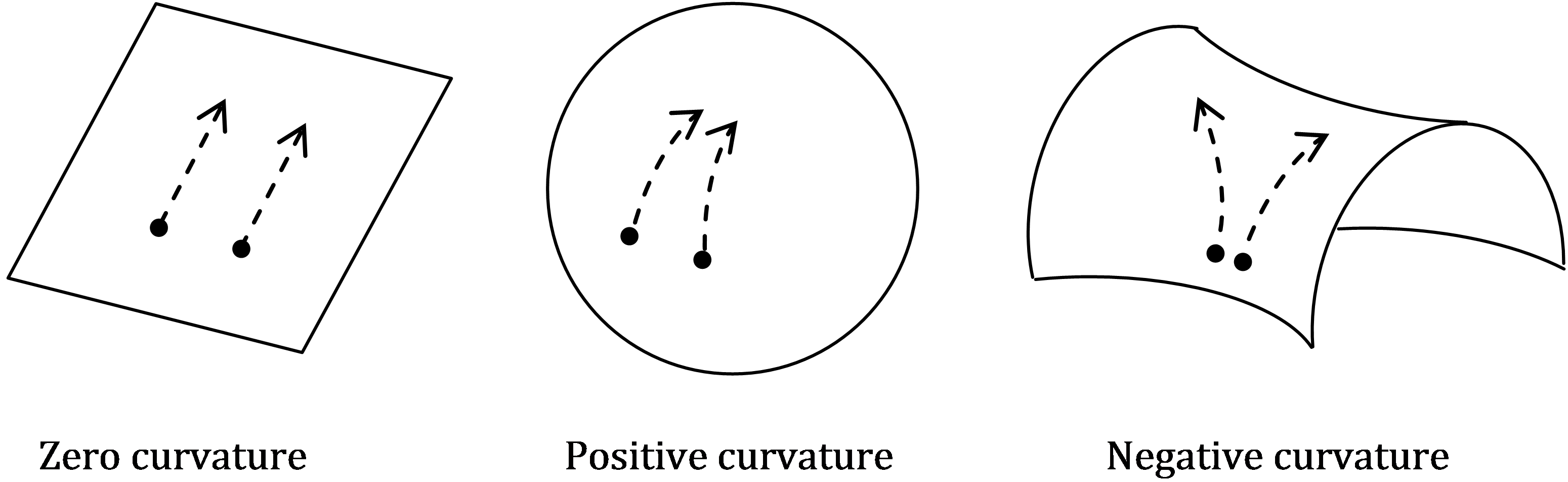}
  \caption{Three types of curvature and the surfaces that best illustrate them.}
\end{figure}

The best example of positive curvature is the surface of a sphere. If we think about this surface in geographical terms, we conclude that the meridians are geodetic, but not the parallel ones, except for the equator, which is a geodetic. In this case, R1 undergoes an important modification, namely, (R2) \textit{if the particles move following initially parallel geodesics, their paths intersect} (see Fig. 3, centre). Following the example of the sheet of paper, this result can be used as a criterion to define the positive curvature.\\

The best example of negative curvature is the surface of a saddle. In this case, it can be shown that, (R3) \textit{two initially parallel geodesics diverge} (Fig. 3, right). We take this result as a criterion to define negative curvature.\\

Now, if we consider a sufficiently small region of the surface of the sphere or the saddle, we discover that the Euclidean geometry is fulfilled with a high degree of approximation, since locally, the curvature is negligible, the geodesics are almost straight and the surface is almost flat. Then, in a very small region on the surface of the sphere or on the saddle, two initially parallel geodesics seem to keep their mutual distance constant.\\

\begin{figure}
  \centering
    \includegraphics[width=0.5\textwidth]{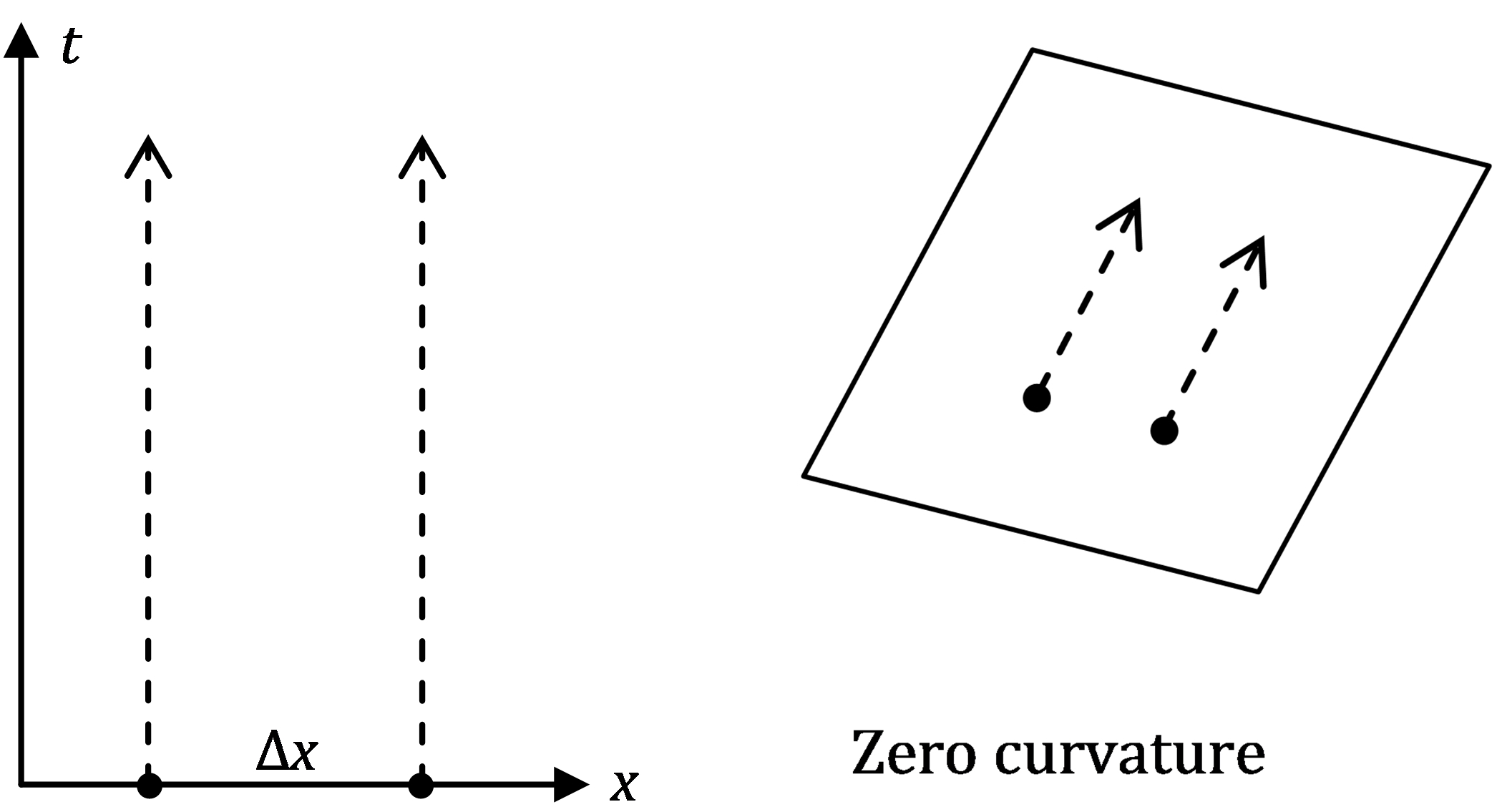}
  \caption{Two particles at rest separated by a distance $\Delta x$ in a region of the universe without gravity, describe parallel and straight geodesics in a spacetime diagram, indicating zero curvature.}
\end{figure}

The analogy between a curved surface and gravity is remarkable. To properly understand this analogy, it is necessary to keep in mind that a particle that moves freely in spacetime, without any force acting on it, will describe a spacetime geodetic. If there is only gravity, the movement of the particles will also be free, since we know that gravity is not a force, but a manifestation of the curvature of spacetime.\\

With the above ideas in mind, imagine two test particles that are left to stand at a distance $\Delta x$. Since by definition we can assume that the gravity between the test particles is zero, the separation $\Delta x$ will remain constant through time, and if we draw their geodetic trajectories on a position $x$ versus time $t$ diagram (spacetime diagram), we will see that they move along straight lines that remain parallel, as shown in Fig. 4. In analogy with the surface of a sheet of paper, we will say that spacetime is flat, and as will be made clear shortly, we can attribute this zero curvature to the absence of gravity.\\

Let us keep the particles at a distance $\Delta x$, but now let us introduce a massive celestial body like Earth. Let the particles free fall toward the centre of the Earth in the radial direction. Under these conditions, the particles will move along geodesics and the distance $\Delta x$ will decrease as time passes. If we extend the paths of the particles to the centre, they intersect (Fig. 5, left). By drawing the geodesics of the particles on a spacetime diagram (Fig. 5, centre), we see that they initially appear parallel, but then begin to approach until they intersect\footnote{In Fig. 5, it is true that $\Delta x \propto t^{n} (n \neq 1)$, which explains why the geodesic trajectory of each particle in the spacetime diagram is curved. For example, for small distances, the approximate relationship is $\Delta x \propto t^{2}$, which implies that in this case the geodesics are parabola segments.}, describing curved paths. This is analogous to what happens on the surface of the sphere. According to Einstein, we can interpret the approach of the particles as a consequence of the positive curvature of spacetime around the Earth in the transverse direction. The effect of gravity is to then curve spacetime.\\

\begin{figure}
  \centering
    \includegraphics[width=0.6\textwidth]{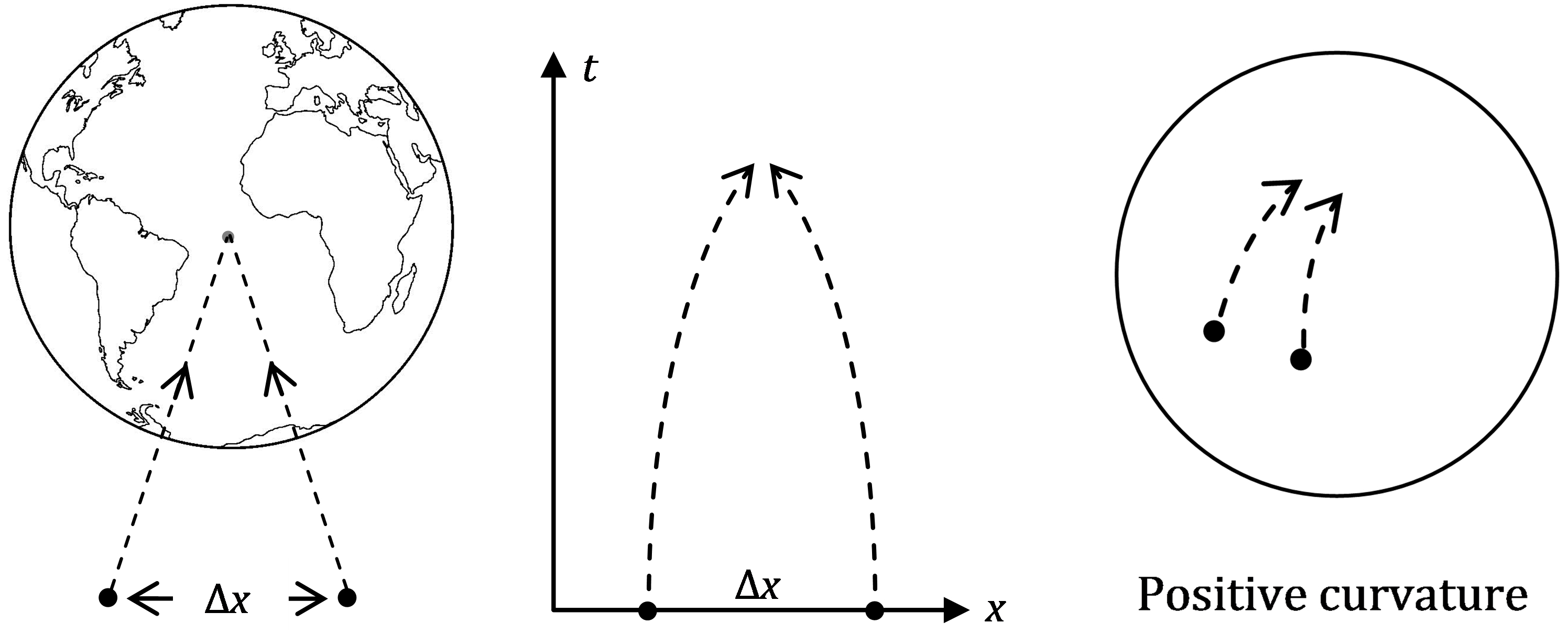}
  \caption{Left: Two particles in free fall toward Earth approach each other in the direction transverse to the radius. Centre: Spacetime diagram for the particles in the left picture, where it is observed that their geodesics are approaching, indicating positive curvature.}
\end{figure}

If we now assume that the particles are located in the same radial direction while free falling, they will move away from each other and the distance $\Delta x$ will increase with time, since the particle closest to Earth will be attracted with more force than the farthest (Fig. 6, left). If we again draw the geodesics of the particles on a spacetime diagram (Fig. 6, centre), we see that they initially appear parallel but then deviate and begin to move away\footnote{In Fig. 6, it is again true that $\Delta x \propto t^{n}$ and therefore the geodetic of each particle is curved.}, describing curved paths. This is analogous to what happens on the surface of the saddle. According to Einstein, we can interpret the separation of the particles as a consequence of the negative curvature of spacetime in the radial direction. Once again, we see that the effect of gravity is to curve spacetime.\\

How are the above conclusions related to the Einstein equivalence principle? Remember that this principle establishes that locally gravity is annulled in a free falling reference frame in a gravitational field. However, a region with zero gravity is a region with zero curvature. This means that the trajectory of an object (like the astronaut in Fig. 2) that is free falling is a locally straight geodesic, where special relativity is locally valid. Then, adding on the locally straight geodesics, we recover the spacetime curvature and therefore the gravity, which is the domain of GR.\\

In summary, the Einstein equivalence principle ensures that GR contains special relativity as a particular case, just as geometry on the surface of the sphere or saddle contains flat Euclid geometry as a particular case.

\begin{figure}
  \centering
    \includegraphics[width=0.6\textwidth]{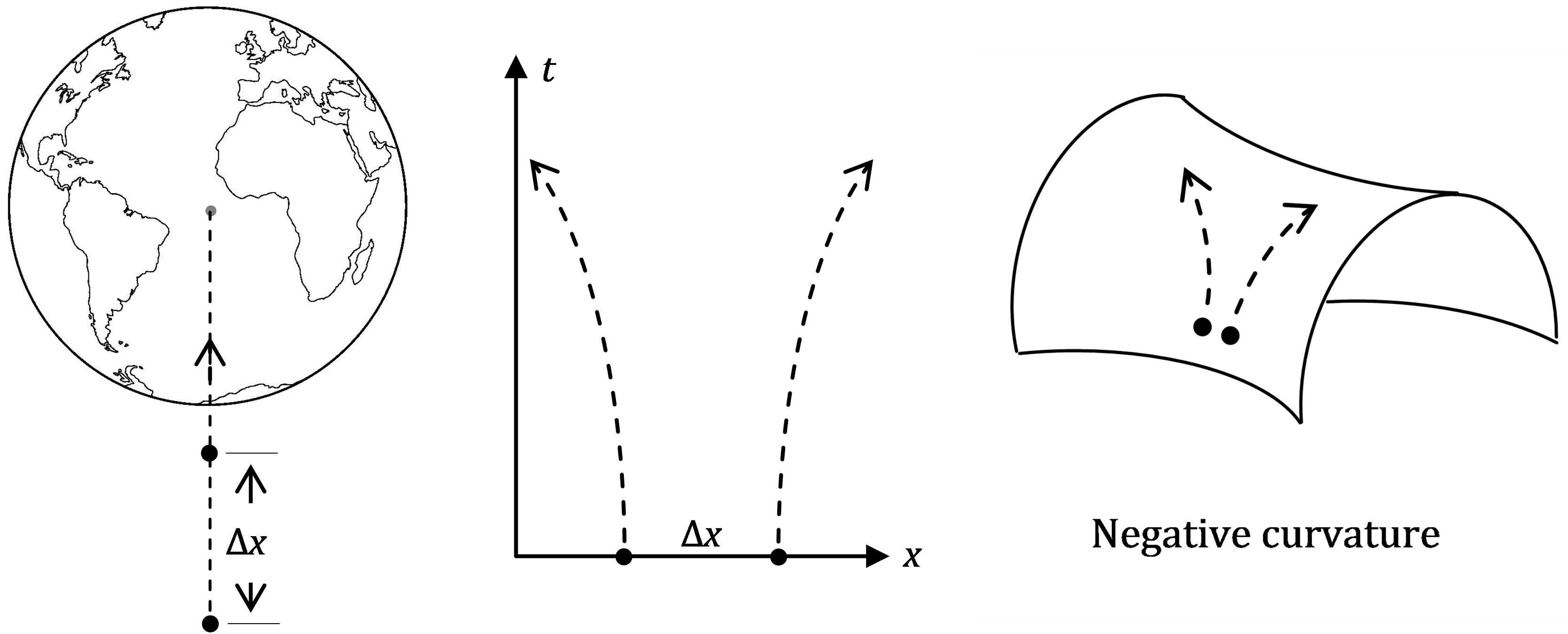}
  \caption{Left: Two particles that free fall in the direction of the Earth’s radius and move away from each other. Centre: Spacetime diagram for the particles in the left picture, where it is observed that their geodesics separate, indicating negative curvature.}
\end{figure}

\section{Final comments}

Perhaps the example of the astronaut and the analogies between gravity and curved surfaces could leave the idea that Einstein's notion of curvature is nothing more than a sophisticated interpretation of the Newtonian concept of gravitational force. If this idea were correct, employing GR or the law of universal gravitation would be a matter of personal preference. However, Newton and Einstein's perspectives are not equivalent, as they lead to different predictions, and one of the main objectives of the classical tests proposed by Einstein was to determine which of the perspectives is correct. In Part II of this article, we will analyse in detail the classical tests and we will discover that the theory that best describes physical reality is not Newton's law of gravitation but GR.

\section*{Acknowledgments}
I would like to thank to Daniela Balieiro for their valuable comments in the writing of this paper. 

\section*{References}

[1] A. Einstein, Die Grundlage der allgemeinen Relativitätstheorie, Annalen der Physik, 354 (1916) 769-822.

\vspace{2mm}

[2] A. Einstein, The Collected Papers of Albert Einstein, Princeton University Press, Princeton, 1997.

\vspace{2mm}

[3] LIGO-Collaboration, Observation of Gravitational Waves from a Binary Black Hole Merger, Phys. Rev. Lett., 116 (2016) 0611021 - 06110216.

\vspace{2mm}

[4] EHT-Collaboration, First M87 Event Horizon Telescope Results. I. The Shadow of the Supermassive Black Hole, The Astrophysical Journal Letters, 875:L1 (2019) 1-17.

\vspace{2mm}

[5] EHT-Collaboration, First M87 Event Horizon Telescope Results. II. Array and Instrumentation, The Astrophysical Journal Letters, 875:L2 (2019) 1-28.

\vspace{2mm}

[6] EHT-Collaboration, First M87 Event Horizon Telescope Results. III. Data Processing and Calibration, The Astrophysical Journal Letters, 875:L3 (2019) 1-32.

\vspace{2mm}

[7] EHT-Collaboration, First M87 Event Horizon Telescope Results. IV. Imaging the Central Supermassive Black Hole, The Astrophysical Journal Letters, 875:L4 (2019) 1-52.

\vspace{2mm}

[8] EHT-Collaboration, First M87 Event Horizon Telescope Results. V. Physical Origin of the Asymmetric Ring, The Astrophysical Journal Letters, 875:L5 (2019) 1-31.

\vspace{2mm}

[9] EHT-Collaboration, First M87 Event Horizon Telescope Results. VI. The Shadow and Mass of the Central Black Hole, The Astrophysical Journal, 875: L6 (2019) 1-44.

\vspace{2mm}

[10] C. Bambi, Introduction to General Relativity A Course for Undergraduate Students of Physics, Springer, Singapore, 2018.

\vspace{2mm}

[11] A. Einstein, Über das Relativitätsprinzip und die aus demselben gezogenen Folgerungen, Jahrbuch der Radioaktivität und Elektronik, 4 (1907) 411-462.

\vspace{2mm}

[12] A. Einstein, Zur Elektrodynamik bewegter Körper, Annalen der Physik, 17 (1905) 891-921.

\vspace{2mm}

[13] K. Krane, Modern Physics, 3 ed., John Wiley and Sons, Hoboken, 2012.

\vspace{2mm}

[14] P.A. Tipler, R.A. Llewellyn, Modern Physics, 6 ed., W. H. Freeman and Company, New York, 2012.

\vspace{2mm}

[15] J. Pinochet, Einstein ring: Weighing a star with light, Phys. Educ., 53 055003 (2018).

\vspace{2mm}

[16] J.A. Wheeler, A journey into gravity and spacetime, W. H. Freeman and Company, New York, 1990.

\end{document}